# The Gluon Propagator[†]

Jeffrey E. Mandula

*U. S. Department of Energy, Washington, DC 20585, U. S. A.*

**Abstract**

We discuss the current state of what is known non-perturbatively about the gluon propagator in QCD, with emphasis on the information coming from lattice simulations. We review the specification of the lattice Landau gauge and the procedure for calculating the gluon propagator on the lattice. We also discuss some of the difficulties in non-perturbative calculations — especially Gribov copy issues. We trace the evolution of lattice simulations over the past dozen years, emphasizing how the improvement in computations has led not only to more precise determinations of the propagator, but has allowed more detailed information about it to be extracted.

## 1   Introduction

In quantum field theory, the Green's functions carry all the information about the theory's physical and mathematical structure. Aside from the vacuum expectation values of fields, the moduli which parametrize the phase structure of a field theory, the two point functions are its most basic quantities. From this point of view, the gluon propagator may be thought of as the most basic quantity of QCD. Even without quarks, in a pure Yang-Mills theory, the gluon propagator is well defined.

At short distances or equivalently large momentum transfers, because of asymptotic freedom we expect that perturbation theory should be sufficient to describe any Green's function. By contrast, at large distances or small momenta, there is no available analytic method to pin down the behavior of the Green's functions. Furthermore, the fact that there are no asymptotic gluon states raises the possibility that the gluon propagator may be quite different than those associated with stable particles. As a result of its centrality from the field theortic perspective, the infrared limit of the gluon propagator has been subject of calculation and speculation since QCD was accepted as the correct theory of the strong interactions in the early 1970's.

---

[†]Talk given at the Dick Slansky Memorial Symposium, Los Alamos, May, 1998

Ideas about the structure of the gluon propagator have been informed by many attractive hypotheses and conceptual puzzles. Among the recurring themes are —

> The relation of the gluon propagator, specifically its infrared behavior, to confinement
> The behavior of Green's functions in a theory in which one expects that none of the quanta of the fundamental fields are physical particles in that theory
> How the Green's functions express such general properties of field theory as spectral positivity
> How the absence of any physical gluon states can be compatible with any non-zero gluon propagator.

Note that despite the absence of physical asymptotic gluon states, gluons are real — every bit as real as quarks. They are also observed in the same way — through the jets of hadrons that result when they are produced in high energy collisions. Their presence was indirectly inferred from the original deep inelastic electron-proton scattering experiments performed at SLAC in the late 1960's and early 1970's. There they were needed to account for the fraction of the momentum of the proton that was not attributable to quarks, the proton's electrically charged constituents. This fraction, according to the so-called momentum sum rule, was close to 50%. Gluon jets were directly observed experimentally in experiments at DESY in the 1980's. The problem of having fundamental constituents that only occur confined inside hadrons is not a physical nor a conceptual one. It is purely a problem of reconciling this physical structure with our notions of how Green's functions behave in field theory.

The plan of this paper is the following: In Section 2, we briefly review the situation in the ultraviolet, where perturbation theory holds, and use this discussion to fix notation. In Section 3, we discuss some of the ideas that have been advanced regarding the behavior of the gluon propagator in the infrared. In Section 4, we discuss the procedure for calculating the gluon propagator on the lattice. We introduce the lattice Landau gauge, and discuss its implementation. We also discuss some of the difficulties in non-perturbative calculations — especially Gribov copy issues. In Section 5 we trace the evolution of lattice simulations over the past dozen years, emphasizing how the improvement in computations has led not only to more precise determinations of the propagator, but has allowed more detailed information about it to be extracted. In Section 6 we summarize what we now know, from non-perturbative lattice studies, about the gluon propagator, and also what is still obscure about it.

## 2 The Gluon Propagator in the Ultraviolet

Since in this discussion we shall be exclusively concerned with the gluon propagator, we restrict our considerations to pure Yang-Mills theory, with no quarks. The gluon propagator is the Fourier transform of the time-ordered matrix element of two gluon fields $A_\mu^a(x)$, where for gauge group $SU(N)$ the index $a$ runs from 1 to $N^2 - 1$.

$$D_{\mu\nu}^{ab}(q) = -i \int d^4x\, e^{iq \cdot x} \langle 0|T(A_\mu^a(x) A_\nu^b(0))|0\rangle \tag{1}$$

In covariant gauges the propagator has the kinematic form

$$D_{\mu\nu}^{ab}(q) = -i\delta^{ab}\left[\left(g_{\mu\nu} - \frac{q_\mu q_\nu}{q^2}\right) D(q^2) + \alpha \frac{q_\mu q_\nu}{q^2} D^L(q^2)\right] \tag{2}$$

The parameter $\alpha$ specifies the gauge, and the Landau gauge is $\alpha = 0$.

To zeroth order in perturbation theory, the propagator is the same as in QED,

$$D_0(q^2) = \frac{1}{q^2} \tag{3}$$

To any finite order in perturbation theory, this power dependence remains valid. It is possible that when all orders are summed, there could result an anomalous dimension

$$D(q^2) \sim \frac{1}{q^{2+\gamma}} \tag{4}$$

What the value of $\gamma$ may be, or even if it is non-zero, is not known.

## 3 The Gluon Propagator in the Infrared

In the infrared, the situation regarding the gluon propagator is much murkier. It is instructive to summarize briefly some of the more thoroughly studied ideas and speculations about the infrared behavior of the gluon propagator:

A) <u>The gluon propagator explicitly displays confinement.</u>

If the confinement potential is linear, it can be expressed in terms of the exchange of quanta whose momentum space propagator behaves like

$$D(q^2) \sim \frac{1}{q^4} \qquad (q^2 \to 0) \tag{5}$$

This behavior was hypothesized by Mandlestam [1] in the late 1970's, and it was made the characteristic of the phenomenological model studied for many years by Baker, Ball, and Zachariasen [2]. More recently, by studying certain truncations of the Schwinger-Dyson equations, this behavior has been advocated by Brown and Pennington [3]. From the point of view that confinement is a result of a sort of "Dual Meissner Effect", the idea that the gluon propagator should express confinement directly is in some sense rather heretical.

B) <u>The propagator has a non-zero anomalous dimension.</u>

Marenzoni, Martinelli, Stella, and Testa [4] carried out lattice simulations of the gluon propagator and interpreted their results, specifically the observation that the propagator fell off quite differently than in perturbation theory, in terms of an anomalous dimension. Their fits to lattice data were consistent with the anomalous dimension being consistent with $\gamma \approx 1$. Their preferred functional form was

$$\frac{1}{Z\left(m^2 + q^2\left(\frac{q^2}{\Lambda^2}\right)^\alpha\right)} \tag{6}$$

C) <u>The propagator acquires an effective mass.</u>

The earliest lattice simulations of the gluon propagator in the Landau gauge, by Mandula and Ogilvie [5] and by Gupta, Guralnik, Kilcup, Patel, Sharpe, and Warnock [6] were interpreted in terms of a massive particle propagator.

$$D(q^2) \sim \frac{1}{q^2 + m^2} \tag{7}$$

D) <u>The propagator vanishes at vanishing momentum.</u>

This behavior, which is often described as the propagator having a pair of poles with conjugate complex masses $m = b\, e^{\pm i\pi/4}$ was hypothesized by Gribov [7], in connection with his study of gauge copies. It has been advocated on a number of different grounds, by Stingl [8], Cudell and Ross [9], Smekal, Hauck, and Alkofer [10], Zwanziger [11], Namislowski [12], and Bernard, Parrinello, and Soni [13]. The gluon propagator in this scenario may take the explicit form in the infrared

$$D(q^2) \sim \frac{q^2}{q^4 + m^4} \qquad (8)$$

This form is an explicit realization of a theorem due to Zwanziger [11], namely that on the lattice, for any finite spacing but in the infinite volume limit, the gluon propagator must vanish at $q^2 = 0$.

$$\lim_{N \to \infty} D(q^2 = 0) = 0 \qquad (9)$$

$N$ is the number of sites on each side of the lattice.

E)  The propagator takes its perturbative form.

The form for the gluon propagator incorporated into all QCD models used in simulations to design experimental detectors and interpret the results of collider experiments is simply the perturbative form

$$D(q^2) = \frac{1}{q^2} \qquad (10)$$

## 4    The Gluon Propagator on the Lattice

Lattice simulations of the gluon propagator have been carried out since the late 1980's, and are still being pursued. The goal of these studies has remained to try to arrive at a definitive understanding of the gluon propagator's infrared behavior. More specifically, one wishes to have as accurate a numerical determination of $D(q^2)$ as one can, over the largest range of momenta, and for the numerical result to converge to an analytic picture of the gluon propagator. One hopes to articulate the character of $D(q^2)$ such that it expresses physics of a theory without physical gluon asymptotic states.

Each of the possibilities listed above is sensibly motivated, in that each incorporates some known or expected property of QCD. As with all lattice calculations, the quality of the results, measured by the statistical precision, the size of the lattice spacing, or the total volume of space-time simulated, have steadily improved over time. In order to appreciate what has and has not been learned, we must first review the proper definitions of operators and gauge conditions, and the sources of errors on the lattice.

## 4.1 The Kinematics of Discretization

On a finite lattice, momentum is a periodic discrete variable. Denoting the lattice spacing by $a$ and the number of sites per side by $N$, each component of the momentum takes the values

$$q_\mu = 0, \pm\frac{2\pi}{aN}, \pm 2\frac{2\pi}{aN}, ..., \pm\frac{\pi}{a} \tag{11}$$

The kinematic range of the dimensionless momentum,

$$aq \equiv \sqrt{\sum_\mu aq_\mu\, aq_\mu} \tag{12}$$

is

$$aq \in [0, 2\pi] \tag{13}$$

The free propagator on the lattice is a periodic function of the lattice momentum and we can use it to define a lattice corrected momentum.

$$D^{(0)}(q) = \frac{1}{\frac{4}{a^2}\sum \sin^2\frac{aq_\mu}{2} + m^2} \equiv \frac{1}{\hat{q}^2 + m^2} \tag{14}$$

The momentum so defined absorbs much of the lattice artifact errors in propagators. It has the kinematic range

$$a\hat{q} \in [0, 4] \tag{15}$$

For small momentum it approaches the ordinary dimensionless momentum.

## 4.2 The Lattice Landau Gauge

On the lattice, where the basic variables are the unitary matrices $U_m(x)$ that express parallel transport, the gauge potential may be defined as [5] [14]

$$A_\mu(x) = \frac{U_\mu(x) - U_\mu^\dagger(x)}{2iag} - Tr\frac{U_\mu(x) - U_\mu^\dagger(x)}{6iag} \tag{16}$$

The lattice Landau gauge condition could be defined in terms of it by

$$\Delta_\mu A_\mu(x) \equiv \sum_\mu A_\mu(x) - A_\mu(x - \hat\mu) = 0 \tag{17}$$

However, it is truer to the continuum situation to formulate it as a maximization condition

$$\underset{g(x)}{Max} \sum_{x,\mu} Re\, Tr\, U_\mu^g(x)$$

$$U_\mu^g(x) \equiv g(x)\, U_\mu(x)\, g(x + \hat\mu)^\dagger \tag{18}$$

The maximization condition implies the finite difference one, and has the virtue that it excludes very unsmooth gauge configurations which would maximize the trace on some sites and minimize it on others, in the extreme case on alternating sites. The maximization condition, in this sense, carries the smoothness character of the continuum Landau gauge.

Any gauge condition expressible as $f(U) = 0$ can be implemented by following the Fade'ev-Popov procedure. This consists of writing the path integral of any quantity $O(U)$ with a gauge invariant measure, multiplying by 1 in the form of a delta function of the gauge condition times the reciprocal of its Jacobian, the Fade'ev-Popov determinant, judiciously interchanging the order of the path integrals, and finally reexpressing the result in terms of an unfixed measure again.

$$\begin{aligned}
\int DU\, e^{-S}\, O(U)\Big|_{f(U)=0} &= \int DU \int Dg\, e^{-S}\, O(U)\, \Delta_{FP}(U)\, \delta(f(U^g)) \\
&= \int Dg \int DU\, e^{-S}\, O(\bar U)\, \Delta_{FP}(U)\, \delta(f(U^g)) \\
&= \int DU\, e^{-S}\, O(\bar U)
\end{aligned} \tag{19}$$

Here, for fixed $g$, $\bar U$ is the gauge transform of $U$ to the $f(U) = 0$ gauge:

$$\bar U[U] = U^g \qquad f(\bar U) = 0 \tag{20}$$

In contrast to perturbation theory, where the evaluation of $\Delta_{FP}$ gives rise to the introduction of ghost fields, in simulations there is no need to compute the Fade'ev-Popov determinant. The correct adjustment to the measure is built into the simulation recipe

Perform the simulation without specification of the gauge, but for each lattice configuration, transform the link variables to the $f(U) = 0$ gauge before evaluating and averaging the path integrand.

*4.3    Gribov Copies*

Gribov copies [7] are a serious conceptual problem in using lattice simulation to understand the behavior of Green's functions in a gauge theory. Their existence on the lattice was investigated starting in the early 1990's [15] [16] [17]. Expressing the gauge condition as a maximization condition at each site avoids some trivial lattice artifact copies, but all the analogues of the continuum Gribov copies are still present. From the first it was clear that the treatment of Gribov copies in a simulation could have a substantial impact on its results for gauge propagators [18]. Because of the uncertainly they lend to simulations of gauge dependent quantities, Gribov copies have been studied by several groups, and studies of their impact on lattice simulations continue [19] [20] [21]. Variant gauge conditions and algorithms continue to be investigated as well [22] [23] [24].

A conceptual procedure for selecting one out of several copies is to take the maximization condition as an absolute maximum. The space of configurations which are absolute maxima is called the fundamental modular domain, and the rule that the path integral should be restricted to the fundamental modular domain removes copies in principal, except on the boundary of the domain [25] [26] [27] [28]. Unfortunately, there is no practical procedure known for actually finding the fundamental modular domain. All methods of gauge fixing are liable to end on copies. Experience in simulations has shown that there are typically many copies, and that quite different configurations can have nearly the same value of the maximization functional.

An inconvenience, though a serious one, is that gauge fixing is a notoriously slow process. Many procedures have been advocated for accelerating the process, and all work fairly well if the size of the lattice is not too large. However, for very large lattices gauge fixing seems to become much slower, and all acceleration methods seem to loose much of their effectiveness.

The situation is in some sense as bad as it can be: the role of Gribov copies is not fully understood, there is no perfect recipe for dealing with them, yet they do seem to matter in that there is ample evidence from simulations that the manner in which they are treated sometimes has a significant impact on the final result of a simulation.

## 5  Results from the Lattice

In this section we will describe the progress that has been made over the dozen or so years that the simulations of the gluon propagator have been carried out.

### 5.1  Earliest Simulations

The first lattice simulations of the gluon propagator were carried out by Mandula and Ogilvie [5] and by Gupta, Guralnik, Kilcup, Patel, Sharpe, and Warnock [6] in 1987. With the computers available at that time, the statistical quality of their signals deteriorated very quickly with lattice time. Therefore they expressed their results in terms of Euclidean lattice time at zero spacial momentum. The graphs of their results are shown in Figs. 1 and 2.

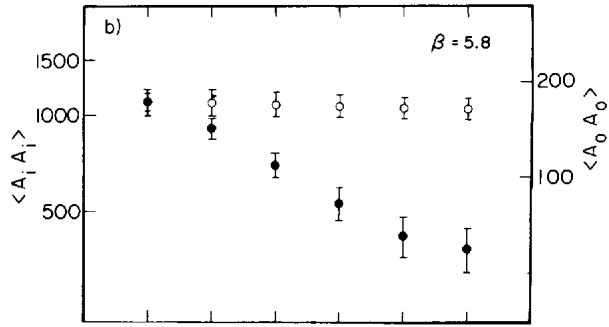

**Figure 1** The zero spatial momentum gluon propagator vs. lattice time, from Mandula and Ogilvie, Ref. [5].

Mandula and Ogilvie's results come from using a $4^3 \times 10$ lattice at $\beta = 5.8$. The open circles are the time-time component of the $D_{\mu\nu}(\vec{q} = \vec{0}, t)$ propagator, which is flat in lattice Landau gauge, while the filled circles are the space-space components, which carry the dynamical information.

The work of Gupta, Guralnik, Kilcup, Patel, Sharpe, and Warnock used the largest lattices that had been employed for lattice gauge simulations to that time, $18^3 \times 42$. They used $\beta = 6.2$ as the lattice coupling, and employed the most powerful supercomputer that was available for lattice simulations, a CRAY at Los Alamos National Lab.

The major conclusion from these analyses follow from the fact that the dynamical components of the propagator seem to fall linearly over an extended range, which, since the plots are on a semi-log scale, is the behavior expected for a massive particle. At the largest distances the lattice periodicity leads to an enhancement, clearly observable from Fig. 2.

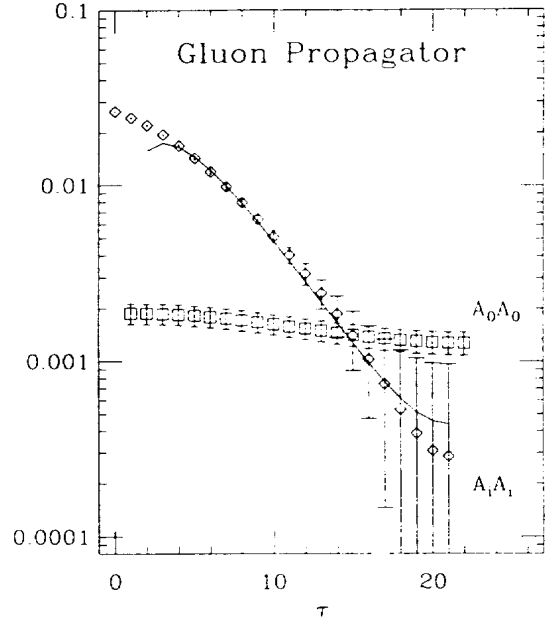

**Figure 2** The zero spatial momentum gluon propagator vs lattice time, from Gupta, Guralnik, Kilcup, Patel, Sharpe, and Warnock, Ref [6].

Another salient feature evident from the figures is that the data clearly follow a curve which is concave downwards for small lattice time. This is quite strange behavior, because it implies that propagator's spectral function is not positive definite. This is easy to see. The curvature of an arbitrary positive linear combination of straight exponential (fixed mass) decays is

$$\frac{d^2}{dt^2} \ln\left(\sum_i c_i e^{-m_i t}\right) = \frac{\left(\sum_i c_i e^{-m_i t}\right)\left(\sum_i c_i m_i^2 e^{-m_i t}\right) - \left(\sum_i c_i m_i e^{-m_i t}\right)^2}{\left(\sum_i c_i e^{-m_i t}\right)^2} \quad (21)$$

For all positive $c_i$, this expression is positive as a consequence of the Schwartz inequality.

### 5.2  *The Effective Gluon Mass*

The time slice to time slice falloff of the gluon propagator provides a natural definition of an effective gluon mass.

$$m_{eff}(t) \equiv -\ln \frac{D(\vec{q}=0, t+1)}{D(\vec{q}=0, t)} \quad (22)$$

For such a definition to be useful requires more computational resources than were available in 1987, when the best that could be done was a global fit to determine the best average mass over the full extent in lattice time.

With considerably greater computational resources, in 1993 Bernard, Parrinello, and Soni [13] and Marenzoni, Martinelli, Stella, and Testa [4] carried out simulations of the gluon propagator with sufficient precision to infer an effective mass value as a function of the lattice time. The results from Ref. [13] are shown in Fig. 3.

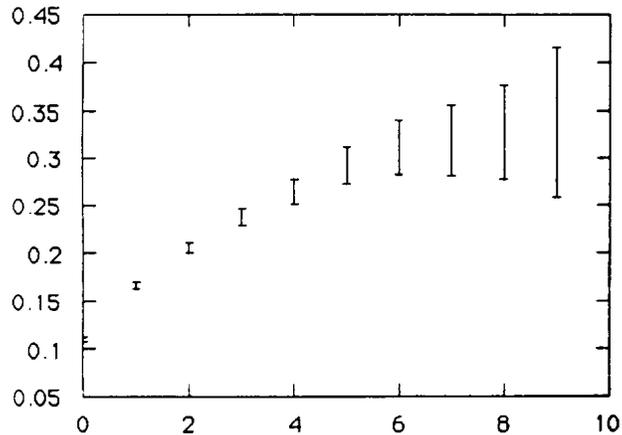

**Figure 3** The effective gluon mass vs. lattice time, from Bernard, Parrinello, and Soni, Ref. [13].

Bernard, Parrinello, and Soni used $16^3 \times 40$ lattices at $\beta = 6.0$. Evidently, the effective mass grows with (Euclidean) time, at least for small times where the lattice data are best. There is no conclusion to be drawn from this analysis about whether or not it levels out to a fixed asymptotic value. This is another demonstration that the propagator is not described by a positive spectral function. For any such, the effective mass would be a monotonically falling function of Euclidean time.

*5.3    The Most Recent Lattice Results*

The state of gluon simulations continues to improve, as much more powerful computers have become available for such studies. Recently, Leinweber, Skullerud, Williams, and Parrinello [29] have carried out a high statistics study simulation on a very large lattice, $32^2 \times 64$ sites, at $\beta = 6.0$. They expressed their results in momentum space, Fig. 4.

The figure gives $a^2 q^2 D(q)$, the momentum-space propagator scaled by $a^2 q^2$, for all values of the lattice 4-momentum. The factor $aq$ is the lattice corrected momentum. The dispersion in the values for moderate $q^2$, which is well outside their statistical errors, indicates that lattice artifacts are still present.

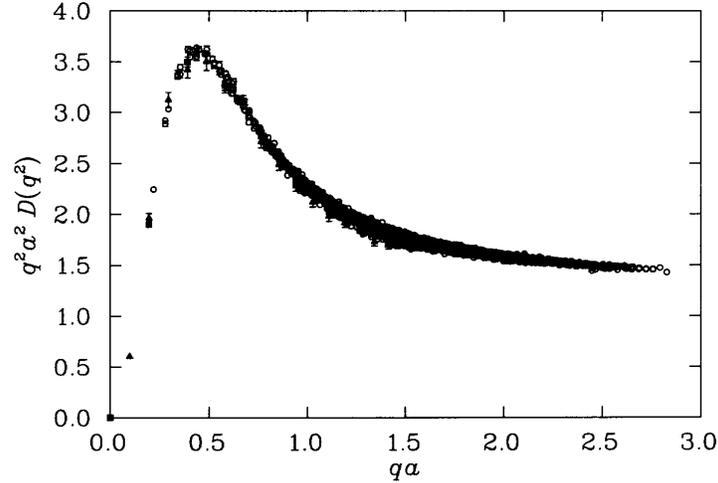

**Figure 4** The momentum space gluon propagator scaled by $q^2$, from Leinweber, Skullerud, Williams, and Parrinello, Ref [29].

This graph displays in yet a different way the fact that the gluon propagator is not describable in terms of a positive spectral function. For a free, massless particle, the graph would be flat, and for a free massive particle, or a general propagator with a positive spectral density, the graph would be monotonically increasing and concave downwards everywhere.

The graph is also incompatible with the Gribov form,

$$D(q^2) = \frac{q^2}{q^4 + m^4} \qquad (23)$$

This would also give a monotonically increasing curve. Finally, if there is an anomalous dimension, it is certainly very small.

## 6  Conclusions: What have we learned?

The essential conceptual problem about the gluon propagator is that there are no asymptotic states associated with it. In a Lehmann-Källen representation, *all* the intermediate states that contribute are non-physical, that is, they lie in gauge-variant sectors of the full Hilbert space of the quantum field theory. It is just the unfamiliarity of this situation that has given rise to suggestions about the analytic structure of the gluon propagator ranging from its being an entire function to its having an essential singularity at the origin.

The most striking observation about the gluon propagator, one that was seen from the earliest simulations, has held up in subsequent analyses. It is that the spectral function describing the gluon propagator is not positive definite. Further simulations have not yet given a definitive picture of the propagator's structure, but they have ruled out some of the plausible suggestions, including some inspired by the earlier simulations.

Simulations of the gluon propagator are no longer compatible with a "complex mass" form à la Gribov. The point is not that one is certain that the gluon propagator is finite at $q^2 = 0$, but rather that it seems that it cannot vanish as rapidly as $q^2$ as $q^2 \to 0$.

The question of an anomalous dimension is still open, although a substantial one cannot be squared with the latest simulations, at least for large $q^2$. One should note though that in those simulations, the propagator falls much more steeply than $1/q^2$ for intermediate values of the momentum.

A final puzzle is Zwanziger's theorem, specifically that even on the largest lattices there is no sign of the vanishing of the propagator at $q^2 = 0$. Here the problem may be analytic rather than computational. The statement of the theorem is that at $q^2 = 0$, all gluon Green's functions vanish as the lattice volume goes to infinity. Even for an infinite lattice, the theorem and its proof give no indication of what the rate of approach to 0 might be. It might be very weak, and might also strongly depend of the total lattice volume.

If the goal of getting a definitive picture of the gluon propagator has not been fully realized as yet, the progress in the past twelve years has been impressively substantial.

**Acknowledgments**

The occasion for this Symposium was the untimely death of Dick Slansky. The author wishes to express his appreciation to Fred Cooper and Geoff West for organizing this meeting to honor Dick's memory. He also wishes to thank them for encouraging this contribution to the Symposium — a review of a subject that Dick supported from its beginnings.